\documentclass[prb,twocolumn,showpacs,floatfix]{revtex4}

\usepackage{graphicx}
\usepackage{amsmath,amssymb}
\usepackage{array}
\usepackage{multirow}

\begin{document}

\title{{\sl Ab initio} study of proper topological ferroelectricity
  in layered perovskite La$_2$Ti$_2$O$_7$}

\author{Jorge L\'opez-P\'erez}
\author{Jorge \'{I}\~{n}iguez}

\affiliation{Institut de Ci\`{e}ncia de Materials de Barcelona (CSIC),
  Campus UAB, 08193 Bellaterra, Spain}

\begin{abstract}
  We present a first-principles investigation of ferroelectricity in
  layered perovskite oxide La$_2$Ti$_2$O$_7$ (LTO), one of the
  compounds with highest Curie temperature known (1770~K). Our
  calculations reveal that LTO's ferroelectric transition results from
  the condensation of two soft modes that have the same symmetry and
  are strongly coupled anharmonically. Further, the leading
  instability mode essentially consists of rotations of the oxygen
  octahedra that are the basic building block of the perovskite
  structure; remarkably, because of the particular topology of the
  lattice, such O$_6$ rotations give raise to a spontaneous
  polarization in LTO. The effects discussed thus constitute an
  example of how nano-structuring -- provided here by the natural
  layering of LTO -- makes it possible to obtain a significant polar
  character in structural distortions that are typically non-polar. We
  discuss the implications of our findings as regards the design of
  novel multifunctional materials. Indeed, the observed proper
  ferroelectricity driven by O$_6$ rotations provides the ideal
  conditions to obtain strong magnetoelectric effects.

\end{abstract}

\pacs{77.84.-s, 61.50.Ah, 75.85.+t, 71.15.Mb}





\maketitle

\section{Introduction}

Because of their physical appeal and technological importance,
ferroelectrics and related materials have been the object of continued
attention for decades.\cite{linesglass,samara01,modernferroelectrics}
Bulk oxides with the ideal perovskite structure -- ranging from
classic ferroelectric BaTiO$_3$ to strong dielectric
Ba$_{1-x}$Sr$_{x}$TiO$_3$, or from piezoelectric
PbZr$_{1-x}$Ti$_{x}$O$_3$ to relaxor
(PbMg$_{1/3}$Nb$_{2/3}$O$_3$)$_{1-x}$-(PbTiO$_3$)$_{x}$ -- have been
especially well studied. Indeed, partly thanks to a number of key
contributions from first-principles theory,\cite{rabe07} we now
understand the fundamental atomistic origin of the ferroelectric (FE)
and response properties of the most important members of this family.
During the past decade, the focus has increasingly shifted towards
nano-structured materials, especially in the form of thin films. Work
on films has led to a better understanding of how elastic (i.e., the
epitaxial strain exerted by a substrate) and electric (e.g., the
imperfect screening of the depolarizing field associated to
particular metallic electrodes) boundary conditions affect the FE
state.\cite{junquera08} Further, it has been shown that exotic FE
properties can be obtained in artificially created super-lattices, as
e.g. in the recently discussed PbTiO$_3$/SrTiO$_3$
heterostructures.\cite{bousquet08}

The emergence of magnetoelectric multiferroics\cite{fiebig05} (i.e.,
materials with coupled magnetic and FE orders) has contributed to
refuel interest in ferroelectrics, especially in what regards {\sl
  unconventional} mechanisms for ferroelectricity. If we restrict
ourselves to oxides with the ideal {\em AB}O$_3$ perovskite structure,
ferroelectricity (usually driven by a {\em B}-site transition metal
with an n$d^{0}$ electronic configuration) and magnetism (which
requires localized $d$ or $f$ electrons) seem to be mutually
exclusive, the known exceptions being very
scarce.\cite{filippetti02,bhattacharjee09} So far, the most notable
ways around this problem consist in (i) having ferroelectricity driven
by the {\em A}-site cation, as in room temperature multiferroic
BiFeO$_3$,\cite{catalan09} and (ii) moving away from the ideal
perovskite structure and resorting to other mechanisms for
ferroelectricity, as e.g. in the improper ferroelectrics
YMnO$_3$\cite{vanaken04,fennie05} and
Ca$_3$Mn$_2$O$_7$.\cite{benedek10}

\begin{figure}
\label{fig1}
\includegraphics[width=0.95\columnwidth]{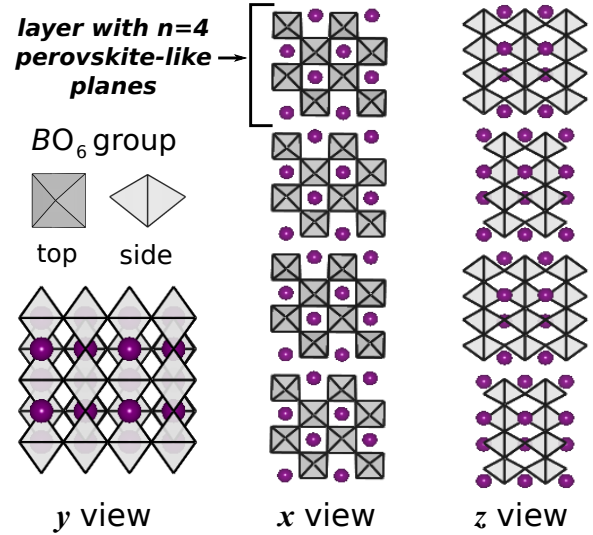}
\caption{(Color on-line.) Different views of the layered perovskite
  structure of the {\em A}$_{n}${\em B}$_{n}$O$_{3n+2}$ compounds for
  $n=$~4, where $n$ is the number of perovskite-like planes within a
  layer. Only the {\em A} atoms (as balls) and {\em B}O$_6$ octahedra
  are shown. The defined Cartesian axes $x$, $y$, and $z$ (which
  follow the convention adopted in
  Ref.~\protect\onlinecite{ishizawa82}) are used throughout the
  paper.}
\end{figure}

This is the context of our work on the layered perovskite oxide
La$_{2}$Ti$_{2}$O$_{7}$ (LTO), whose structure is sketched in Fig.~1.
LTO is one of the highest-temperature ferroelectrics known, with a
Curie point ($T_{\rm C}$) of approximately 1770~K,\cite{nanamatsu74}
and is lately being considered for applications as a high-$T$
piezoelectric.\cite{yan09,bruyer10} The FE transition temperature of
LTO is enormous, especially when compared with the $T_{\rm C}$'s of
cubic perovskite titanates one might have expected to be similar to
LTO: most significantly, BaTiO$_3$ (BTO) becomes ferroelectric at
about 400~K, and the related compound SrTiO$_3$ remains (a quantum)
paraelectric down to 0~K. In order to explain the surprising behavior
of LTO, we need to identify the atomistic origin of its FE
instability. LTO might owe its very high $T_{\rm C}$ to the kind of
mechanisms known to operate in the cubic titanates (in essence,
long-range dipole-dipole interactions that destabilize the non-polar
phase\cite{posternak94,ghosez99}), which might somehow be enhanced by
the peculiar topology of the LTO lattice. Alternatively, LTO might
present some new form of very strong FE instability. Either way, the
study of LTO will provide us with information that might be useful for
the design of new ferroelectrics.

Let us also note that LTO is a member of the family of oxides with
general formula La$_{n}$Ti$_{n}$O$_{3n+2}$,\cite{lichtenberg01} whose
structure can be seen as the ideal cubic perovskite periodically
truncated along the [011]$_{\rm c}$ direction of the cubic lattice,
$n$ being the number of perovskite-like planes within one layer (see
Fig.~1). These structures are thus related to the well-known
Ruddlesden-Popper, Aurivillius, and Dion-Jacobson families of layered
perovskites,\cite{rao98} for which the truncation direction is
[001]$_{\rm c}$ and which include very famous members such as
(La,Ba)$_2$CuO$_4$, the parent compound of high-$T_{\rm C}$
superconductors. The basic features of the electronic structure of the
La$_{n}$Ti$_{n}$O$_{3n+2}$ compounds are controlled by $n$: The La and
O atoms can be assumed to be in their most common ionization states --
i.e., La$^{3+}$ and O$^{2-}$ --, which implies that the Ti cations
will present a positive charge of $3+4/n$. Accordingly, the number of
3$d$ electrons in the Ti atoms will be $1-4/n$, which allows us to
move quasi-continuously from the Ti-3$d^{0}$ configuration of the
$n$=4 compound (that is our La$_2$Ti$_2$O$_7$, the family member with
smallest $n$ reported in the literature) to the Ti-3$d^{1}$
configuration of the $n$=$\infty$ compound (which has the prototype
perovskite structure). The variety of electronic phenomena observed
for intermediate values of $n$ -- including phases of semi-conducting,
normal metallic, and low-dimensional metallic
character\cite{lichtenberg01} -- constitutes an additional motivation
to study in detail the structural behavior of the relatively simple
end member La$_2$Ti$_2$O$_7$.

The paper is organized as follows. In Section~II we describe the
theoretical approach and first-principles methods used in this
work. We present and discuss our results in Section~III, which is
split in the following way. In Section~III.A we describe our results
for the structure of the high-temperature paraelectric (PE) and FE
phases of LTO. In Section~III.B we show that the PE phase presents a
strong FE instability whose {\em topological} nature is discussed in
detail. In Section~III.C we describe the phase transition between the
PE and FE phases, and discuss the energetics of the transformation. In
Section~III.D we present our results for the spontaneous polarization,
dielectric, and piezoelectric properties. Having established LTO's
peculiarity, which is made manifest by means of a detailed comparison
with prototype compound BTO, in Section~III.E we show that LTO also
presents some features that are clearly reminiscent of the usual FE
oxides with the ideal perovskite structure. In Section~III.F we
outline the exciting implications that our results have for the design
of novel magnetoelectric materials. Finally, in Section~IV we
summarize and present our conclusions. Whenever it is possible, we
compare our first-principles results with the (scarce) experimental
information available for LTO.

\section{Methodology}

\subsection{Theoretical approach to La$_2$Ti$_2$O$_7$}

We adopted the usual first-principles approach to the investigation of
structural phase transitions of the displacive type, which is
routinely applied with great success to FE perovskite
oxides.\cite{kingsmith94} Here we describe a simplified version of
such an approach that is sufficient for our study of LTO (i.e., we
will assume one-dimensional instability modes, will not discuss how to
deal with cell strains, etc.). In essence, one needs to identify a
reference equilibrium phase of high symmetry (HS) -- which corresponds
to the high-temperature PE phase of the compound -- and study its
stability against all possible structural distortions. More precisely,
one writes the energy of the crystal as the following Taylor series:
\begin{equation} E \; = \; E^{0} \, + \, \frac{1}{2} \sum_{m,n} K_{mn}
u_m u_n \, + {\cal O}(u^3) \, , \end{equation}
where $E^0$ is the energy of the HS phase. The $u_m$ variables
represent the structural distortions of the reference
configuration; for the study of a FE transition, one can typically
restrict oneself to the distortions compatible with the reference unit
cell, i.e., those associated to the $\Gamma$-point of the Brillouin
zone of the PE phase. The so-called {\sl force-constant matrix}
$\boldsymbol{K}$ is the central quantity one needs to compute, as its
negative eigenvalues (if any) correspond to unstable structural
distortions -- i.e., {\sl soft} modes -- that may result in a phase
transition. Let us use the term {\em mode stiffness} to refer to the
eigenvalues of $\boldsymbol{K}$, which we denote by $\kappa_{s}$ with
$s$ running from 1 to the dimension of the force-constant matrix. Note
that the magnitude of a negative $\kappa_{s}$ determines the {\sl
  strength} of the structural instability, and thus the likelihood of
observing it experimentally. (In general, one may find several,
possibly competing, instabilities of a HS phase; it is by no means
guaranteed that all of them will lead to experimentally observable
phase transitions.) Once a soft mode is identified, one can readily
study the corresponding low-symmetry (LS) phase -- by distorting the
crystal according to the soft mode eigenvector and relaxing the
resulting structure -- and the energetics of the instability -- by
computing the usual double-well potential connecting the HS and LS
phases.

In order to apply this program to LTO, we had to tackle one
fundamental difficulty: We could not find any experimental information
on the structure of the high-temperature PE phase of this
compound.\cite{fn:highTphase} Note that this is never a problem when
one works with the usual FE perovskites, where the ideal cubic
perovskite structure is the reference phase of choice. Such a choice
is obviously correct in cases like those of BTO or PbTiO$_3$, where
the cubic PE phase is experimentally accessible for $T>T_{C}$; more
remarkably, this choice is also the most physically sound one for
materials (e.g., BiFeO$_3$) whose cubic PE phase is not easy to access
experimentally, as the samples tend to melt before reaching the
corresponding transition temperature. In the general case, the problem
of choosing an appropriate HS reference phase for the theoretical
study of displacive phase transitions has long been solved. The basic
idea is to look for {\sl pseudo-symmetries} (i.e., {\sl slightly}
broken symmetries) of the known LS structure; we can thus identify
possible HS phases that would transform into the known LS structure
upon a relatively small distortion. This is a very powerful strategy
that can lead to the discovery of complex phase transition sequences
(e.g., when more than one possible HS phases are found) and, in
particular, has been used to identify previously unnoticed FE
transitions.\cite{abrahams71,kroumova02}

\begin{figure}
\label{fig2}

\includegraphics[width=0.95\columnwidth]{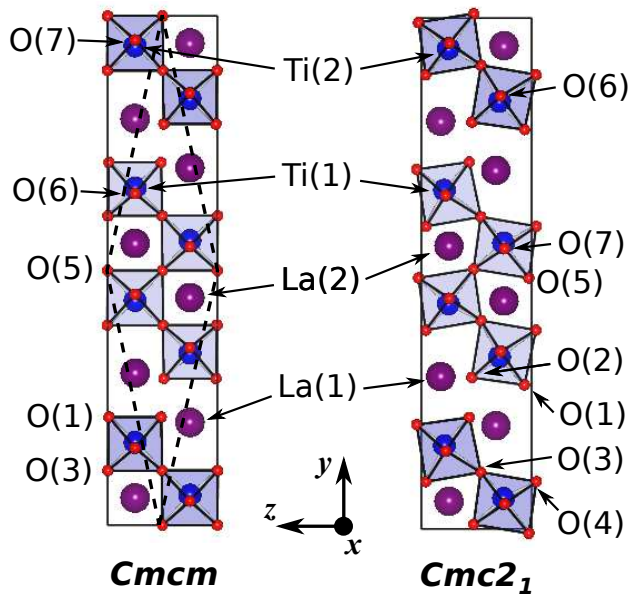}

\caption{(Color on-line.) Structures of the $Cmc2_1$ (FE) and $Cmcm$
  (PE) phases of La$_2$Ti$_2$O$_7$ studied in this work. The small
  (red), medium (blue), and big (violet) balls represent O, Ti, and La
  atoms, respectively. The shaded polyhedra are top-viewed O$_6$
  groups. The two phases have the same 44-atom conventional cell
  depicted in the figure; the 22-atom unit cell, which is also common
  to both, is indicted with dashed lines in the $Cmcm$ case. The
  defined Cartesian axes $x$, $y$, and $z$ (which follow the
  convention adopted in Ref.~\protect\onlinecite{ishizawa82}) are used
  throughout the paper. The symmetry-independent atoms are labeled as
  in Table~I; note that oxygens O(1) and O(3) of the $Cmcm$ structure
  split, respectively, into O(1)/O(2) and O(3)/O(4) in the $Cmc2_1$
  structure.}

\end{figure}

Using widely available crystallographic tools,\cite{kroumova01} we
applied the pseudo-symmetry analysis to LTO and obtained a very clear
prediction: We found that the FE phase of this compound, which
presents space group $Cmc2_1$ and a 22-atom primitive unit cell, is
most likely associated to a PE phase with the same primitive cell and
space group $Cmcm$ (see sketch in Fig.~2). We thus performed our
first-principles study using this $Cmcm$ phase as our HS reference
structure.

A few additional points are in order: (1) The pseudo-symmetry analysis
resulted in relatively large atomic displacements connecting the HS
and LS phases, with maximum values of about 0.4~\AA\ corresponding to
the La atoms. Note that the magnitude of such distortions is expected
to reflect the associated HS-LS transition
temperature,\cite{abrahams68} which is indeed very high in this
case. (2) Experimental studies of the compounds Sr$_2$Nb$_2$O$_7$ and
Ca$_2$Nb$_2$O$_7$,\cite{nanamatsu71,brandon70} which share with LTO
the same layered structure and a similar (n$d^{0}$) electronic
configuration of the transition metal atoms, show that they present a
high-temperature phase with $Cmcm$ space group. (3) The choice of
$Cmcm$ as our PE space group results in specific predictions for the
symmetry of the soft modes that would be associated with a FE
transition to the $Cmc2_{1}$ phase. Indeed, the leading instability
should transform with the $B_{1u}$ irreducible representation of
$mmm$, the point group of the $Cmcm$ phase. As we will see, our
first-principles results confirmed these expectations, thus supporting
our choice of space group for the PE phase.

\subsection{Details of the calculations}

We used the local density approximation (LDA) to density functional
theory (DFT) as implemented in the first-principles package {\sc
  VASP}.\cite{vasp} To represent the ionic cores we used the
projector-augmented wave scheme,\cite{vasp-paw} solving explicitly for
the following electrons: La's 5$p$, 5$d$, and 6$s$; Ti's 3$p$, 3$d$,
and 4$s$; and O's 2$s$ and 2$p$. The electronic wave functions were
represented in a plane-wave basis truncated at 400~eV. We always
worked with the 44-atom cell of LTO sketched in Fig.~2 (which is the
conventional cell for both $Cmcm$ and $Cmc2_1$ phases), and used a
6$\times$1$\times$5 $k$-point grid for Brillouin zone integrations. We
checked these calculation conditions were well-converged by monitoring
the computed equilibrium structure, bulk modulus, and phonon
frequencies of the $Cmcm$ phase. For the calculation of the
force-constant matrix $\boldsymbol{K}$ and the dielectric and
piezoelectric responses, we employed a simple finite-displacement
scheme and the corresponding linear-response tensor formulas, which
can be found e.g. in Ref.~\onlinecite{wu05}. We only computed the
lattice-mediated part of the static dielectric response, which has
been repeatedly shown to dominate the effect in ferroelectric oxides.

Our BTO simulations were analogous to the above described ones. We
solved explicitly for Ba's 5$s$, 5$p$, and 6$s$ electrons (treating Ti
and O as above), and used a 400~eV cut-off for the plane-wave
basis. We worked with the 5-atom unit cell of BTO, and used a
6$\times$6$\times$6 $k$-point grid for Brillouin zone integrations.

\section{Results and Discussion}

\subsection{Structure of the $Cmcm$ and $Cmc2{_1}$ phases}

Table~I shows our results for the equilibrium structure of the $Cmcm$
(PE) and $Cmc2_{1}$ (FE) phases of LTO considered in this study. For
$Cmc2_{1}$ we also report the experimental structure measured at
1173~K by Ishizawa {\sl et al}.\cite{ishizawa82} using X-ray
diffraction.

\begin{table}[t!]
\setlength{\extrarowheight}{1mm}

\label{table1}

\caption{Computed equilibrium structures
    of the $Cmcm$ and $Cmc2_1$ phases of La$_2$Ti$_2$O$_7$ discussed
    in the text. We show in parenthesis the experimental values
    reported in Ref.~\protect\onlinecite{ishizawa82} for the $Cmc2_1$
    phase.}

\vskip 1mm

\begin{tabular*}{0.95\columnwidth}{@{\extracolsep{\fill}}ll}
\hline\hline
$Cmcm$          & $a$~=~3.891~\AA \;\; $b$~=~25.720~\AA \;\; $c$~=~5.465~\AA \\
                        & $\alpha$~=~$\beta$~=~$\gamma$~=~90$^{\circ}$ \\  [1ex]
\hline
\end{tabular*}
\begin{tabular*}{0.95\columnwidth}{@{\extracolsep{\fill}}ccccc}
Atom & Wyc. & $x$    & $y$    & $z$    \\
La(1)   & 4c   & 0 & 0.2924 & 1/4 \\
La(2)   & 4c   & 0 & 0.4453 & 3/4 \\
Ti(1)    & 4c   & 1/2 & 0.3389 & 3/4 \\
Ti(2)    & 4c   & 1/2 & 0.4437 & 1/4 \\
O(1)    & 8f    & 1/2 & 0.2881 & 0.9931 \\ 
O(3)    & 8f    & 1/2 & 0.3982 & 0.9869 \\ 
O(5)    & 4a   & 1/2 & 1/2 & 1/2 \\ 
O(6)    & 4c   & 0 & 0.3458 & 3/4 \\ 
O(7)    & 4c   & 0 & 0.4525 & 1/4 \\ [1ex]
\end{tabular*}

\begin{tabular*}{0.95\columnwidth}{@{\extracolsep{\fill}}ll}
\hline\hline
$Cmc2_{1}$          & $a$~=~3.845~\AA \;\; $b$~=~25.626~\AA \;\;
$c$~=~5.464~\AA \\
         & ($a$~=~3.954~\AA \;\; $b$~=~25.952~\AA \;\;
$c$~=~5.607~\AA) \\
         & $\alpha$~=~$\beta$~=~$\gamma$~=~90$^{\circ}$ \\  [1ex]
\hline
\end{tabular*}
\begin{tabular*}{0.95\columnwidth}{@{\extracolsep{\fill}}ccccc}
Atom & Wyc. & $x$    & $y$    & $z$    \\
La(1)   & 4a  & 0 & 0.2978 & 0.1675 \\
            &      &    & (0.2981) & (0.1757) \\
La(2)   & 4a   & 0 & 0.4464 & 0.7515 \\
            &      &    & (0.4461) & (0.7500) \\
Ti(1)    & 4a   & 1/2 & 0.3369 & 0.7069 \\
            &      &    & (0.3370) & (0.7095) \\
Ti(2)    & 4a   & 1/2 & 0.4407 & 0.2422 \\
            &      &    & (0.4404) & (0.2452) \\
O(1)    & 4a   & 1/2 & 0.2803 & 0.9245 \\ 
            &      &    & (0.2818) & (0.9350) \\
O(2)    & 4a   & 1/2 & 0.2960 & 0.4399 \\ 
            &      &    & (0.2964) & (0.4580) \\
O(3)    & 4a   & 1/2 & 0.3843 & 0.0415 \\ 
            &      &    & (0.3891) & (0.0390) \\
O(4)    & 4a   & 1/2 & 0.4072 & 0.5549 \\ 
            &      &    & (0.4077) & (0.5420) \\
O(5)    & 4a   & 1/2 & 0.4905 & 0.9724 \\ 
            &      &    & (0.4912) & (0.9750) \\
O(6)    & 4a   & 0 & 0.3460 & 0.7426 \\ 
            &      &    & (0.3472) & (0.7200) \\
O(7)    & 4a   & 0 & 0.4507 & 0.2604 \\ 
            &      &    & (0.4511) & (0.2550) \\ [1ex]
\hline\hline
\end{tabular*}

\end{table}

While the overall agreement between the experimental and theoretical
FE structures is acceptable, the deviations affecting some atomic
positions and lattice constants are larger than what is usual for this
type of calculations. For example, the predicted position of the La(1)
atom differs notably from the experimentally determined one; more
significantly, the computed $a$ and $c$ lattice constants deviate from
the experimental values by almost a 3\%. We attribute these
discrepancies to the fact that we are comparing our first-principles
results, which correspond to the limit of 0~K, with structural data
taken at very high temperatures. Given that thermal expansion and
other temperature-driven effects are not included in our simulations,
our results seem compatible with the experimental information.

\subsection{Ferroelectric instabilities of the $Cmcm$ phase}

As described in Section~II.A, we studied the structural stability of
the $Cmcm$ phase against $\Gamma$-point distortions (compatible with
the PE unit cell) by computing the corresponding force-constant matrix
$\boldsymbol{K}$. From the diagonalization of $\boldsymbol{K}$ we
obtained two negative eigenvalues that correspond to two structural
instabilities. The computed mode stiffnesses are $-$0.81~eV/\AA$^{2}$
and $-$0.01~eV/\AA$^{2}$, respectively. Hence, according to our
calculations, LTO's $Cmcm$ phase presents a structural instability
comparable in strength with the FE soft mode of BTO's PE phase (for
which we obtained $-$2.74~eV/\AA$^{2}$), as well as a marginally
unstable mode with nearly zero stiffness.

Both instabilities transform with the $B_{1u}$ irreducible
representation of the $mmm$ point group of the PE phase; more
precisely, they are infra-red active (polar) modes that involve the
development of a polarization along the $z$ direction defined in
Fig.~2. Such $B_{1u}$ modes break the $C_{2y}$, $C_{2x}$, $I$, and
$\sigma_{z}$ point symmetries of the PE phase (where $I$ is the
spatial inversion, $C_{2\alpha}$ stands for a two-fold rotation around
axis $\alpha$, and $\sigma_{\alpha}$ is a mirror plane perpendicular
to direction $\alpha$), leading to the $Cmc2_{1}$ space group. Hence,
the obtained $B_{1u}$ soft modes are exactly the kind of instabilities
that can drive a ferroelectric phase transition between the $Cmcm$ and
$Cmc2_{1}$ phases. Our first-principles results thus support the
correctness of our working hypothesis, i.e., that the studied $Cmcm$
structure is indeed the high-symmetry PE phase of LTO.

We can gain insight into the origin of ferroelectricity in LTO by
inspecting the eigenvector of the strongest instability mode, denoted
by $\boldsymbol{\xi}_{1}$ in the following. In essence,
$\boldsymbol{\xi}_{1}$ involves a rotation of the O$_6$ oxygen
octahedra around the $x$ axis, as sketched in Fig.~3(a). Hence, the
displacements of the equatorial oxygens amount to most of the
eigenvector (11\% of the norm of $\boldsymbol{\xi}_{1}$ is associated
to O(1) displacements, 51\% to O(3), and 28\% to O(5), using the
labels defined in Fig.~2 for the oxygen atoms); there are also
significant La displacements (9\% of the norm), while the Ti atoms
and apical oxygens O(6) and O(7) have a negligible participation in
$\boldsymbol{\xi}_{1}$. The character of the second soft mode
($\boldsymbol{\xi}_{2}$) is more complex: It is dominated by the
displacements of oxygens O(1) and O(5) (with 21\% and 35\% of the
norm, respectively) and involves a significant deformation of the
O$_6$ octahedra; it also presents a large participation of the La
atoms (about 24\% of the norm).

\begin{figure}
\label{fig3}
\includegraphics[width=0.95\columnwidth]{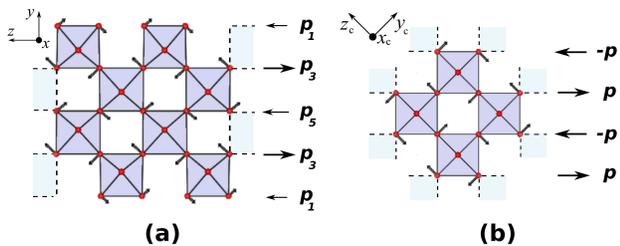}
\caption{(Color on-line.) Panel~(a): sketch of the largest atomic
  displacements associated to the strongest instability mode
  ($\boldsymbol{\xi}_1$) obtained for the $Cmcm$ phase of
  La$_2$Ti$_2$O$_7$. We show the displacements corresponding to one
  layer composed of $n=$~4 perovskite-like planes. The arrows on the
  side represent the electric dipoles associated to the displacement
  of oxygens in different $y$-planes (see text). The dipoles are
  labeled as the oxygen atoms in Fig.~2. Panel~(b): sketch of a
  typical anti-ferrodistortive mode occurring in an ideal
  (non-layered) perovskite structure.}
\end{figure}

These FE instabilities are very different from the ones that are usual
among {\sl AB}O$_3$ oxides with the ideal perovskite structure. The
inset of Fig.~4(b) shows the representative case of BTO: the Ti cation
moves away from the center of the O$_6$ octahedron, which is only
slightly distorted, giving raise to a large electric dipole. Further,
the FE instability in BTO and related materials is known to originate
from the strong interactions between such local
dipoles.\cite{ghosez99} In contrast, the dominant FE instability found
in LTO consists of O$_6$ octahedra rotations, the off-centering of the
Ti atoms being negligible. As quantified in Section~III.D, such a
pattern of atomic displacements does not lead to a large local dipole,
which suggests that the mechanisms responsible for the
ferroelectricity in LTO have to be of a different nature.

Interestingly, structural instabilities involving O$_6$ octahedra
rotations are very common among {\sl AB}O$_3$ perovskites, and are
usually termed anti-ferrodistortive (AFD). Indeed, AFD modes as the
one sketched in Fig.~3(b) are the driving force for most of the
structural phase transitions occurring in these compounds, the
examples including crystals as well-known as SrTiO$_3$, {\em
  Ln}MnO$_3$ and {\em Ln}NiO$_3$ (where {\em Ln} is a lanthanide),
multiferroics Bi{\em M}O$_3$ (where {\em M} is a 3$d$ transition
metal), etc. Accordingly, there is extensive literature devoted to the
study and classification AFD modes in {\sl AB}O$_3$
perovskites,\cite{glazer72,howard04} and it is known that size (e.g.,
the incompatibility of the ionic radii of the {\em A} and {\em B}
cations to form a cubic perovskite lattice) and chemical (as in the
Bi-based compounds that display the so-called {\em stereochemical
  activity}) effects are usually responsible for the occurrence of AFD
distortions.\cite{dieguez-sub} Hence, it is not a surprise to find
that layered perovskite structures may present AFD-like instabilities
that allow for a better fulfillment of steric and/or chemical
constraints at a local level. It is not our goal here to discuss the
occurrence and origin of such soft modes in LTO-like layered
perovskites; such a study should probably involve consideration of a
number of representative crystals (e.g., a few members of the {\em
  A}$_{n}${\em B}$_{n}$O$_{3n+2}$ family) and falls beyond the scope
of this work. Suffice it to say that the leading FE instability that
we found in LTO is essentially analogous to the AFD distortions that
are ubiquitous among perovskite oxides.

However, we must note a critical difference between LTO's AFD-like
mode depicted in Fig.~3(a) and the AFD modes occurring in ideal
perovskite structures [Fig.~3(b)]: the former causes a spontaneous
polarization, while the latter do not. To better explain this, in
Fig.~3(a) we indicate with arrows the electric dipoles that appear as
a consequence of the displacement of the oxygen atoms following the
$\boldsymbol{\xi}_1$ eigenvector. The oxygens in each $y$-oriented
plane give raise to a dipole along the $z$ direction. Using the
definitions in Fig.~3(a), the total dipole associated to one layer
would be $\boldsymbol{p}^{\rm
  layer}=\boldsymbol{p}_{5}+2\boldsymbol{p}_{3}+2\boldsymbol{p}_{1}$.
Since there is no symmetry relationship between the displacements of
the O(1), O(3), and O(5) oxygens in the $\boldsymbol{\xi}_1$
eigenvector, it follows immediately that $\boldsymbol{p}^{\rm layer}$
will be different from zero. Moreover, even if oxygens of different
types were to displace by the same amount, so that
$\boldsymbol{p}_{5}=-\boldsymbol{p}_{3}=\boldsymbol{p}_{1}=\boldsymbol{p}$,
we would still have $\boldsymbol{p}^{\rm layer} = \boldsymbol{p} \neq
0$, as the number of oxygen planes in each layer is odd. Finally, note
that LTO's unstable mode $\boldsymbol{\xi}_1$ involves an identical
distortion of all layers in the structure, and thus gives raise to a
net macroscopic polarization.

On the other hand, AFD distortions of the ideal (non-layered)
perovskite structure do not result in a macroscopic polarization. In
Fig.~3(b) we show a pattern of O$_6$ rotations around the indicated
$x_{\rm c}$ axis; the situation is very similar to the one depicted in
Fig.~3(a), except that in this case the network of O$_6$ octahedra is
not truncated. Again, we indicate with arrows the electric dipoles
that appear as a consequence of the oxygen displacements: oxygens
within a $[011]_{\rm c}$-oriented plane give raise to a dipole
$\boldsymbol{p}$ along $[01\bar{1}]_{\rm c}$. It is apparent that the
addition of all such dipoles gives no net polarization in this case, a
result that can be viewed as a consequence of the three-dimensional
nature of the O$_6$ network.

In conclusion, we have found that the strongest structural instability
in LTO's high-symmetry phase is a very common and simple one: it
involves concerted rotations of the O$_6$ octahedra, much alike the
AFD modes responsible for the structural phase transitions in most
{\em AB}O$_3$ crystals with the ideal perovskite structure. AFD
distortions in the usual perovskites are well-known to be non-polar, a
feature that can be viewed as a consequence of the symmetry and
three-dimensional character of the lattice. In contrast, because the
structure of LTO is split in layers comprising an even number of
perovskite planes, the O$_6$-rotation mode gives raise to a net
polarization in this case. Hence, since the occurrence of a
spontaneous polarization in LTO relies on the layered topology of the
lattice, it seems appropriate to describe this compound as a {\em
  topological ferroelectric}. Let us stress that LTO's peculiar
lattice topology does not seem essential for the structural
instability to exist, but it is critical for it to have a polar
character.

\subsection{Nature of the $Cmcm$-to-$Cmc2_1$ transition}

In addition to the already discussed topological nature of the leading
FE instability, the transformation between LTO's $Cmcm$ and $Cmc2_1$
phases presents a number of interesting aspects that we discuss in the
following. Most importantly, our results show that the transition is
driven by the combined action of the two soft modes discussed in the
previous Section, and suggest that such a cooperation is critical for
it to occur at a very high temperature.

\begin{table}
\setlength{\extrarowheight}{1mm}

\label{table2}
\caption{Energy difference between the FE and PE phases of La$_2$Ti$_2$O$_7$
  and BaTiO$_3$. The FE phases are considered under several elastic
  constraints (see text), which we denote ``fully-relaxed'' (no constraint),
  ``PE volume'', and ``PE cell''. For both LTO and BTO, the energies are
  normalized to the volume of the PE phase so that a
  comparison between compounds can be made. Results given in meV/\AA$^3$.}

\vskip 1mm

\begin{tabular*}{0.75\columnwidth}{@{\extracolsep{\fill}}ccc}
\hline\hline
 & La$_2$Ti$_2$O$_7$ & BaTiO$_3$ \\
\hline
fully-relaxed & $-$1.097 & $-$0.110 \\
PE volume  & $-$1.042 & $-$0.099 \\
PE cell  & $-$0.938 & $-$0.051 \\
\hline\hline
\end{tabular*}

\end{table}

To study a structural phase transition quantitatively, one can start
by comparing the energies of the phases involved. Table~II shows our
results for LTO, along with the analogous data for the $Pm\bar{3}m$
(PE) and $P4mm$ (FE) phases of BTO. Note that the energy change {\sl
  per} unit volume involved in LTO's transition is about one order of
magnitude greater than the corresponding one for BTO. Such an enormous
difference is compatible with the experimentally measured Curie
temperatures, which are 1770~K and 400~K for LTO and BTO,
respectively. Hence, our first-principles results for the energetics
of LTO's FE transition seem consistent with experiments.

Table~II also contains information about the elastic deformation that
accompanies the FE transitions. The properties of FE oxides with the
ideal perovskite structure are known to be strongly sensitive to cell
strains. This is clearly reflected in the results for BTO: If the
compound is forced to keep the cubic cell of the PE structure, the
energy gain involved in the FE transition is reduced by half (i.e., it
drops from $-$0.110~meV/\AA$^3$ to $-$0.051~meV/\AA$^3$). On the other
hand, if we only impose the PE volume (so that the FE cell is allowed
to deform and acquire a $c/a\ne 1$ ratio), the energetics of the
transformation is not strongly affected. In contrast, the analogous
elastic constraints have a relatively small effect in the case of LTO
(i.e., the energy gain drops from $-$1.097~meV/\AA$^3$ to
$-$0.938~meV/\AA$^3$ when we impose the PE cell). Such a weak coupling
between strain and the FE distortion is probably reflecting the
AFD-like character of the instability;\cite{fn:strainAFD} this result
also suggests that LTO will display relatively small piezoelectric
effects as compared with regular FE perovskites.

Let us now consider the atomic displacements that characterize the
$Cmcm$-to-$Cmc2_1$ transformation. In most materials undergoing
displacive transitions, the atomic distortion connecting the HS and LS
phases is essentially captured by the eigenvector of the instability
mode that triggers the transformation. One can easily quantify this by
constructing a {\em distortion vector} $\Delta\boldsymbol{X}$
comprising all the atomic displacements associated to the HS-to-LS
transition, and expressing it in the basis formed by the eigenvectors
$\boldsymbol{\xi}_s$ of the force-constant matrix of the HS
phase. (For simplicity, we constructed our distortion vectors
$\Delta\boldsymbol{X}$ using the atomic positions of a LS phase that
is forced to have the same unit cell as the HS phase.) For BTO, this
analysis led us to the expected result: the soft mode of the PE phase
captures 98\% of the atomic distortion involved in the
$Pm\bar{3}m$-to-$P4mm$ transition. However, the result for LTO was
qualitatively different: the strongest instability mode
($\boldsymbol{\xi}_1$, with $\kappa_1 = -$0.81~eV/\AA$^{2}$) captures
81\% of the total distortion, and the second instability mode
($\boldsymbol{\xi}_2$, with $\kappa_2 = -$0.01~eV/\AA$^{2}$)
contributes with a 15\%. (For both compounds, no other mode
contributes more than a 1\%.) Hence, our results suggest that the two
soft modes $\boldsymbol{\xi}_1$ and $\boldsymbol{\xi}_2$ play an
important role in LTO's FE phase transition.

\begin{figure}
\includegraphics[width=0.95\columnwidth]{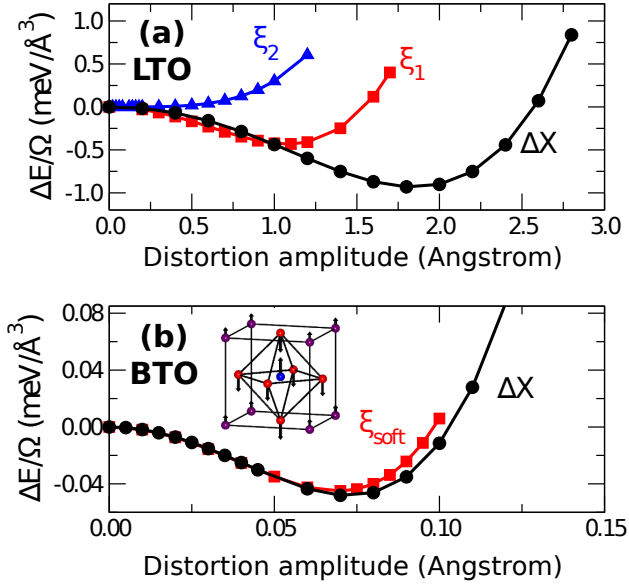}
\caption{(Color on-line.) Panel~(a): Variation of the energy of
  La$_2$Ti$_2$O$_7$ as the $Cmcm$ phase is distorted according to
  several atomic displacement patterns, namely, those corresponding to
  the soft modes $\boldsymbol{\xi}_1$ and $\boldsymbol{\xi}_2$, as
  well as the distortion $\Delta\boldsymbol{X}$ that connects the
  $Cmcm$ and $Cmc2_1$ phases. The 44-atom cell of the $Cmcm$ phase is
  kept fixed in all calculations. Panel~(b): Same as panel~(a) for the
  PE ($Pm\bar{3}m$) to FE ($P4mm$) transition of BaTiO$_3$. The inset
  shows the atomic displacements associated to the single FE soft mode
  of BTO (Ba atoms at the corners of the cubic cell; Ti atom at the
  center of the O$_6$ octahedron). For both LTO and BTO, the energies
  are normalized to the volume of the PE phase so that a comparison
  between compounds can be made.}
\label{fig4}
\end{figure}

Such a large contribution of $\boldsymbol{\xi}_2$ to the structural
transformation seems incompatible with the computed mode stiffnesses;
indeed, the values of $\kappa_1$ and $\kappa_2$ would suggest that
$\boldsymbol{\xi}_2$ is about 80 times weaker than
$\boldsymbol{\xi}_1$ as an instability. To resolve this apparent
contradiction, we computed how LTO's energy changes when the $Cmcm$
phase is distorted according to the individual $\boldsymbol{\xi}_1$
and $\boldsymbol{\xi}_2$ modes. As shown in Fig.~4(a), we found a
large energy reduction associated with the $\boldsymbol{\xi}_1$
distortion, while the $\boldsymbol{\xi}_2$ instability is almost
negligible. We also computed the energy variation as the total
distortion $\Delta\boldsymbol{X}$ is frozen in. Remarkably, as
compared with the results for $\boldsymbol{\xi}_1$, the
$\Delta\boldsymbol{X}$ curve in Fig.~4(a) presents a much deeper
minimum corresponding to a much larger distortion amplitude. This is a
new indication that the $\boldsymbol{\xi}_1$ soft mode cannot explain
LTO's FE transition by itself. This result contrasts with the
situation for BTO [Fig.~4(b)], where the FE soft mode captures the
energetics of the structural transformation almost exactly.

The results of Fig.~4(a) suggest that there is a strong and {\em
  cooperative} coupling between the two instability modes of LTO. To
better describe this effect, let us write the energy of the crystal as
the following Taylor series:
\begin{equation}
\begin{split}
E = & \, E^{Cmcm} + \frac{1}{2} \kappa_1 u_1^2 + \frac{1}{2} \kappa_2
u_2^2  + \frac{1}{4} \alpha_1 u_1^4 + \frac{1}{4} \alpha_2 u_2^4 \\
& + \gamma' u_1^3 u_2+ \gamma'' u_1^2 u_2^2 +
\gamma''' u_1 u_2^3 + {\cal O}(u^6) \, , 
\end{split}
\label{eq:E}
\end{equation}
where $E^{Cmcm}$ is the energy of the $Cmcm$ phase, and $u_1$ and
$u_2$ are, respectively, the amplitudes (in Angstrom) of the
$\boldsymbol{\xi}_1$ and $\boldsymbol{\xi}_2$ distortions. Note that
this expression for the energy is greatly simplified by symmetry, and
that the quadratic parameters coincide with the mode stiffnesses. We
have included in Eq.~(\ref{eq:E}) only the lowest-order couplings
between $u_1$ and $u_2$, which are quantified by the primed $\gamma$
parameters. By fitting to the $\boldsymbol{\xi}_1$ and
$\boldsymbol{\xi}_2$ energy curves of Fig.~4(a), we got $\kappa_1
=-$0.79~eV/\AA$^{2}$ and $\kappa_2 =$~0.00~eV/\AA$^{2}$, in fair
agreement with the values obtained from the diagonalization of
$\boldsymbol{K}$; we also got $\alpha_1 =$~0.65~eV/\AA$^{4}$ and
$\alpha_2 =$~0.66~eV/\AA$^{4}$. Then, to fit the
$\Delta\boldsymbol{X}$ curve we considered distortions characterized
by $u_{1}/u_{2}=0.81/0.15$, as it corresponds to the $Cmc2_1$
phase. Note that, in this case, the fitted quartic coefficient is a
combination of the $\alpha$ and $\gamma$ parameters of
Eq.~(\ref{eq:E}). Given the large $u_{1}/u_{2}$ ratio associated with
the $\Delta\boldsymbol{X}$ distortion, and in view of further tests
discussed in Section~III.D, it seems reasonable to assume that
$\gamma'$ dominates over $\gamma''$ and $\gamma'''$; we thus got
$\gamma' =$~$-$0.35~eV/\AA$^4$. As compared with the computed $\alpha$
coefficents, this clearly is a very strong anharmonic coupling that
favors the combined $\boldsymbol{\xi}_{1} + \boldsymbol{\xi}_{2}$
distortion.\cite{fn:unbounded}

To the best of our knowledge, such a cooperation between soft modes is
rare among perovskite oxides. There are many examples of materials in
which several strong instabilities exist; in most cases, the strongest
one leads to a phase transition that tends to suppress, partially or
totally, the other instabilities. The competition between the FE and
AFD soft modes in compounds like SrTiO$_3$ is a representative and
well studied case.\cite{zhong95} In contrast, we find that the
reciprocal enhancement of the two soft FE modes is critical to explain
the structural transformation in LTO. The $\boldsymbol{\xi}_1$ mode is
clearly the leading instability, and it would occur even in absence of
$\boldsymbol{\xi}_2$. Yet, as the results in Fig.~4(a) show, the
magnitude and strength of the transformation are boosted by the
interaction between $\boldsymbol{\xi}_1$ and
$\boldsymbol{\xi}_2$. Thus, our results seem to suggest that LTO owes
its very high $T_{\rm C}$ to such an interaction; indeed, in view of
Fig.~4(a), it is hard to imagine LTO's $T_{\rm C}$ would remain
essentially the same if the coupling between $\boldsymbol{\xi}_{1}$
and $\boldsymbol{\xi}_{2}$ was suppressed.

Studying from first-principles the temperature dependence of the
$\boldsymbol{\xi}_1$ and $\boldsymbol{\xi}_2$ distortions, and the
details of how these two soft modes freeze in at LTO's FE transition,
is a challenging endeavor that remains for future work. Nevertheless,
a few observations can be made based on the present results. In
principle, one could try to approach the problem by introducing a
Landau potential of the form:
\begin{equation}
\begin{split}
  F - F_0 = \, & \frac{1}{2} A_1 (T-T_{\rm C}) Q_1^2 + \frac{1}{4} B_1
  Q_1^4 + \\ 
&  \frac{1}{2} A_2 Q_2^2 + \frac{1}{4} B_2 Q_2^4 + \\
& C'  Q_1^3  Q_2 + C'' Q_1^2  Q_2^2 + C''' Q_1 Q_2^3  \, , 
\end{split}
\label{eq:F}
\end{equation}
which is written in terms of two one-dimensional order parameters,
$Q_1$ and $Q_2$, that have the same symmetry. In the following
heuristic argument we will consider only the $Q_1^3 Q_2$ crossed term
-- in accordance with our above conjecture regarding the couplings in
Eq.~(\ref{eq:E}), and because this is the most relevant crossed term in
vicinity of the phase transition--,\cite{fn:c} thus assuming that $C''
= C''' =$~0.

If we were dealing with a simple transition, we would have a primary
order parameter $Q_1$ corresponding to the unstable eigenmode of the
$\boldsymbol{K}$ matrix of the high-symmetry phase; the temperature
dependence of the Landau potential would be restricted to the $Q_1^2$
term, as indicated in Eq.~(\ref{eq:F}), and we would have positive
$A_1$ and $B_1$ coefficients. Further, a secondary order parameter
$Q_2$ would be stable by itself, with positive $A_2$ and $B_2$
coefficients. Hence, the corresponding Landau theory would predict a
phase transition at $T=T_{\rm C}$, with
\begin{equation}
Q_1 = \left[ \frac{A_1}{B_1} (T_{\rm C}-T) \right]^{1/2}
\end{equation}
below the transition temperature. (The indicated formulas were derived
assuming that we remain close to the transition
temperature.)  Then, $Q_2$ would present the following temperature
dependence below $T_{\rm C}$:
\begin{equation}
Q_2 = - \frac{C'}{A_2} Q_1^3 = - \frac{C'}{A_2} \left[
  \frac{A_1}{B_1} (T_{\rm C}-T) \right]^{3/2} \, .
\end{equation}
In a simple case, the coefficients in Eq.~(\ref{eq:E}) may be a good
approximation to the coefficients in Eq.~(\ref{eq:F}) in the limit of
low temperatures. Thus, by supplementing the Eq.~(\ref{eq:E}) deduced
from first-principles with a piece experimental information -- i.e.,
the value of the transition temperature $T_{\rm C}$ --, we could
construct the corresponding Landau potential and obtain a quantitative
description of the temperature dependence of $Q_1$ and $Q_2$.

The doubts quickly appear when one tries to apply this model to
LTO. It seems natural to associate $Q_1$ and $Q_2$ with our computed
FE soft modes $\boldsymbol{\xi}_1$ and $\boldsymbol{\xi}_2$,
respectively. However, that identification implies that the transition
temperature for $Q_1$ is independent of $Q_2$; such an approximation
would be an awkward one, given the large influence that
$\boldsymbol{\xi}_2$ has in the energetics of the $Cmcm$-to-$Cmc2_1$
transformation. Additionally, we would be forced to speculate
regarding the $T$-dependence of the $A_2$ coefficient, as our {\em
  guess} for this parameter -- i.e., the $\kappa_2$ of
Eq.~(\ref{eq:E}) which was found to be essentially zero -- does not
comply with the usual requirements for the quadratic coefficient of a
secondary mode. Note that, for example, if we had a small value of
$A_2$ and thus a dominant $B_2$ term, the behavior with temperature of
both $Q_1$ and $Q_2$ would vary: $Q_1$ would have the same functional
$T$-dependence but with a different prefactor, and $Q_2$ would go as
$(T_{\rm C}-T)^{1/2}$ instead of $(T_{\rm C}-T)^{3/2}$. The results of
this study do not allow us to resolve such details of LTO's
transition.

\begin{table}
\setlength{\extrarowheight}{1mm}

\label{table3}
\caption{Non-zero components of the spontaneous polarization ($P^{\rm S}_{z}$,
  given in C/m$^2$), lattice-mediated dielectric tensor
  ($\epsilon_{\alpha\beta}$), and piezoelectric tensor ($d_{\alpha l}$, where
  $l$ labels strain components in Voig notation, given in pC/N) of the
  $Cmc2_1$ phase of La$_2$Ti$_2$O$_7$. For the dielectric tensor, the
  clamped-cell response is given in parenthesis.\protect\cite{wu05} For the
  piezoelectric tensor, the clamped-ion response is given in parenthesis. We
  also show experimental\protect\cite{nanamatsu74} and
  theoretical\protect\cite{bruyer10} values for the $P2_{1}$ phase of LTO that
  is stable at room temperature ($T_{\rm room}$). Note that the results in
  Refs.~\protect\onlinecite{nanamatsu74} and \onlinecite{bruyer10}
  have been adapted to our choice of Cartesian axes.}

\vskip 1mm

\begin{tabular*}{0.9\columnwidth}{@{\extracolsep{\fill}}cccc}
  \hline\hline
  & $Cmc2_1$ phase & \multicolumn{2}{c}{$P2_{1}$ phase} \\
  \cline{2-2}
  \cline{3-4}
  & this work & exp. ($T_{\rm room}$)\protect\cite{nanamatsu74} &
  theory\protect\cite{bruyer10} \\ 
\cline{2-2}\cline{3-3}\cline{4-4}
$P^{\rm S}_{z}$ & 0.29 & 0.05 & 0.08 \\
$\epsilon_{xx}$ & 62 (61) & 52 & - \\
$\epsilon_{yy}$ & 44 (44) & 42 & - \\
$\epsilon_{zz}$ & 65 (54) & 62 & - \\
$d_{z1}$  & 12 (0) & 3 & - \\
$d_{z2}$  &  4 (1) & 6 & - \\
$d_{z3}$  & $-$22 (0) & 16 & - \\
$d_{x5}$  & $-$2 (0) & - & - \\
$d_{y4}$  & 1 (0) & - & - \\
  \hline\hline
\end{tabular*}
\end{table}

\subsection{Polarization and response properties}

Table~III shows the computed spontaneous polarization,
lattice-mediated dielectric tensor, and piezoelectric tensor for the
$Cmc2_1$ phase of LTO. For comparison, we also show results from the
literature corresponding to the $P2_{1}$ phase that is stable at room
temperature.

We obtained a value of 0.29~C/m$^2$ for the spontaneous polarization,
which is comparable with the result of 0.38~C/m$^2$ that we obtained
for the tetragonal phase of BTO. The dielectric response is also very
significant, as the obtained values are comparable with those typical
of FE perovskites (e.g., we got $\epsilon_{zz} =$~23 for the
$z$-polarized tetragonal phase of BTO). As regards piezoelectricity,
the computed responses are considerable but not particularly large;
for example, for the low-temperature rhombohedral phase of BTO, the
piezoelectric coefficients reach values of 200~pC/N,\cite{wu05} while
for LTO we got maximum values of about 20~pC/N. LTO's relatively small
piezoelectric response seems compatible with the minor role that the
cell strains play in determining the energetics of the
$Cmcm$-to-$Cmc2_1$ phase transition, as mentioned in the discussion of
Table~II.

The large spontaneous polarization obtained may seem incompatible with
the AFD-like character of the structural instability
($\boldsymbol{\xi}_1$) that dominates the $Cmcm$-to-$Cmc2_1$
transition. To clarify this point, we performed alternative
calculations using linear-response expressions\cite{wu05} based on the
Born effective-charge tensors $\boldsymbol{Z}^{*}_{i}$ that quantify
the polarization change associated to the displacement of an
individual atom $i$.

The polarization reported in Table~III was obtained in the standard
way: We used the Berry-phase theory of King-Smith and
Vanderbilt\cite{kingsmith93} to compute the variation of
$\boldsymbol{P}$ as the $Cmc2_1$ structure is deformed into a
symmetry-equivalent one with opposite polar distortion; then,
$\boldsymbol{P}^{\rm S}$ is half of the computed polarization
change. $\boldsymbol{P}^{\rm S}$ can also be estimated using the
approximate formula
\begin{equation}
  P^{\rm S}_{\alpha} \approx \sum_{i\beta}
  Z^{*}_{i,\alpha\beta} \Delta X_{i\beta} \, ,
\end{equation}
where $\Delta\boldsymbol{X}$ is the vector capturing the distortion
that connects the $Cmcm$ and $Cmc2_1$ phases, $i$ labels the atoms in
the unit cell, and $\alpha$ and $\beta$ label spatial
directions. Using different choices for the effective-charge tensors
(i.e., those computed for the $Cmcm$ phase, as well as the
corresponding results for the $Cmc2_1$ phase subject to different cell
constraints) we obtained values of $P^{\rm S}_{z}$ in the
0.25-0.32~C/m$^2$ range, which are perfectly compatible with the
result in Table~III.

\begin{table}
\setlength{\extrarowheight}{1mm}

\label{table4}
\caption{Computed effective-charge tensors (given in units of elementary charge)
  for the $Cmcm$ (PE) and $Cmc2_1$ (FE) phases of La$_2$Ti$_2$O$_7$. To
  facilitate the comparison between phases, we list the
  tensors corresponding to all the symmetry-inequivalent atoms of the
  FE phase. Atoms are labeled as in Table~I.}

\vskip 1mm

\begin{tabular*}{0.95\columnwidth}{@{\extracolsep{\fill}}ccc}
\hline\hline
& $Cmcm$ (PE) & $Cmc2_1$ (FE) \\
\cline{2-2}\cline{3-3}
La(1) &
 $\left( \begin{array}{ccc}
4.63 & 0 & 0 \\
0 & 4.17 & 0 \\
0 & 0 & 4.72 
\end{array} \right) $ &
 $\left( \begin{array}{ccc}
4.64 & 0 & 0 \\
0 & 4.09 & $-$0.21 \\
0 & $-$0.69 & 4.71 
\end{array} \right) $ \\
La(2) &
 $\left( \begin{array}{ccc}
4.37 & 0 & 0 \\
0 & 3.76 & 0 \\
0 & 0 & 4.28 
\end{array} \right) $ &
 $\left( \begin{array}{ccc}
4.63 & 0 & 0 \\
0 & 4.13 & 0.41 \\
0 & 0.24 & 4.24 
\end{array} \right) $ \\
Ti(1) &
 $\left( \begin{array}{ccc}
6.91 & 0 & 0 \\
0 & 6.70 & 0 \\
0 & 0 & 5.72 
\end{array} \right) $ &
 $\left( \begin{array}{ccc}
6.16 & 0 & 0 \\
0 & 5.60 & $-$0.36 \\
0 & 0.24 & 5.31 
\end{array} \right) $ \\
Ti(2) &
 $\left( \begin{array}{ccc}
6.33 & 0 & 0 \\
0 & 5.32 & 0 \\
0 & 0 & 7.45 
\end{array} \right) $ &
 $\left( \begin{array}{ccc}
6.11 & 0 & 0 \\
0 & 5.29 & 0.09 \\
0 & $-$0.15 & 6.32 
\end{array} \right) $ \\
O(1) &
 $\left( \begin{array}{ccc}
$-$2.52 & 0 & 0 \\
0 & $-$3.20 & 0.73 \\
0 & 0.86 & $-$3.26 
\end{array} \right) $ &
 $\left( \begin{array}{ccc}
$-$2.52 & 0 & 0 \\
0 & $-$3.14 & 0.53 \\
0 & 0.38 & $-$2.92 
\end{array} \right) $ \\
O(2) &
 $\left( \begin{array}{ccc}
$-$2.52 & 0 & 0 \\
0 & $-$3.20 & $-$0.73 \\
0 & $-$0.86 & $-$3.26 
\end{array} \right) $ &
 $\left( \begin{array}{ccc}
$-$2.15 & 0 & 0 \\
0 & $-$2.35 & $-$0.89 \\
0 & $-$0.90 & $-$3.44 
\end{array} \right) $ \\
O(3) &
$\left( \begin{array}{ccc}
$-$2.08 & 0 & 0 \\
0 & $-$3.27 & $-$1.72 \\
0 & $-$1.56 & $-$3.98 
\end{array} \right) $ &
 $\left( \begin{array}{ccc}
$-$1.76 & 0 & 0 \\
0 & $-$3.33 & $-$1.36 \\
0 & $-$1.38 & $-$2.95
\end{array} \right) $ \\
O(4) &
$\left( \begin{array}{ccc}
$-$2.08 & 0 & 0 \\
0 & $-$3.27 & 1.72 \\
0 & 1.56 & $-$3.98 
\end{array} \right) $ &
 $\left( \begin{array}{ccc}
$-$2.49 & 0 & 0 \\
0 & $-$3.19 & 1.28 \\
0 & 1.19 & $-$3.64 
\end{array} \right) $ \\
O(5) &
 $\left( \begin{array}{ccc}
$-$2.49 & 0 & 0 \\
0 & $-$3.15 & 1.19 \\
0 & 0.98 & $-$3.64 
\end{array} \right) $ &
 $\left( \begin{array}{ccc}
$-$2.63 & 0 & 0 \\
0 & $-$3.11 & 1.04 \\
0 & 1.00 & $-$3.36
\end{array} \right) $ \\
O(6) &
 $\left( \begin{array}{ccc}
$-$5.50 & 0 & 0 \\
0 & $-$2.15 & 0 \\
0 & 0 & $-$1.66
\end{array} \right) $ &
 $\left( \begin{array}{ccc}
$-$5.04 & 0 & 0 \\
0 & $-$2.08 & 0.24 \\
0 & 0.14 & $-$1.75 
\end{array} \right) $ \\
O(7) &
 $\left( \begin{array}{ccc}
$-$5.10 & 0 & 0 \\
0 & $-$1.73 & 0 \\
0 & 0 & $-$2.41 
\end{array} \right) $ &
 $\left( \begin{array}{ccc}
$-$4.99 & 0 & 0 \\
0 & $-$1.91 & $-$0.03 \\
0 & $-$0.02 & $-$2.20 
\end{array} \right) $ \\
\hline\hline
\end{tabular*}
\end{table}

\begin{table}
\setlength{\extrarowheight}{1mm}

\label{table5}
\caption{Computed effective-charge tensors (given in units of elementary charge)
  for the $Pm\bar{3}m$ (PE) and $P4mm$ (FE) phases of BaTiO$_3$. The tensors
  are given in the conventional Cartesian axes for a rectangular
  lattice. They correspond to the following atoms
  (PE phase positions given in relative units): Ba located at (0,0,0),
  Ti at (1/2,1/2,1/2), O(1) at (1/2,1/2,0), and O(2) at
  (0,1/2,1/2). O(1) and O(2) are symmetry-equivalent in the PE phase,
  but not in the $z$-polarized FE phase.}

\vskip 1mm

\begin{tabular*}{0.95\columnwidth}{@{\extracolsep{\fill}}ccc}
\hline\hline
& $Pm\bar{3}m$ (PE) & $P4mm$ (FE) \\
\cline{2-2}
\cline{3-3}
Ba &
 $\left( \begin{array}{ccc}
2.72 & 0 & 0 \\
0 & 2.72 & 0 \\
0 & 0 & 2.72 
\end{array} \right) $ &
 $\left( \begin{array}{ccc}
2.71 & 0 & 0 \\
0 & 2.71 & 0 \\
0 & 0 & 2.81 
\end{array} \right) $ \\
Ti &
 $\left( \begin{array}{ccc}
7.31 & 0 & 0 \\
0 & 7.31 & 0 \\
0 & 0 & 7.31 
\end{array} \right) $ &
 $\left( \begin{array}{ccc}
7.05 & 0 & 0 \\
0 & 7.05 & 0 \\
0 & 0 & 5.83 
\end{array} \right) $ \\
O(1) &
 $\left( \begin{array}{ccc}
$-$2.13 & 0 & 0 \\
0 & $-$2.13 & 0 \\
0 & 0 & $-$5.77
\end{array} \right) $ &
 $\left( \begin{array}{ccc}
$-$1.98 & 0 & 0 \\
0 & $-$1.98 & 0 \\
0 & 0 & $-$4.79
\end{array} \right) $ \\
O(2) &
 $\left( \begin{array}{ccc}
$-$5.77 & 0 & 0 \\
0 & $-$2.13 & 0 \\
0 & 0 & $-$2.13
\end{array} \right) $ &
 $\left( \begin{array}{ccc}
$-$5.62 & 0 & 0 \\
0 & $-$2.14 & 0 \\
0 & 0 & $-$1.96
\end{array} \right) $ \\

\hline\hline
\end{tabular*}
\end{table}

Then, we used the effective-charge tensors of the $Cmcm$ phase (given in
Table~IV) to compute the polarization change associated with the condensation
of the $\boldsymbol{\xi}_1$ and $\boldsymbol{\xi}_2$ soft modes that dominate
the $Cmcm$-to-$Cmc2_1$ transformation. The polar character of the modes is
quantified in terms of the {\em mode effective charges}:
\begin{equation}
  \bar{Z}_{s,\alpha} = \sum_{i\beta} Z^{*}_{i,\alpha\beta}
  \xi_{s,i\beta} \, .
\end{equation}
We obtained $\bar{Z}_{1,z}=$~1.8$e$ and $\bar{Z}_{2,z}=$~12.0$e$,
where $e$ is the elemental charge. These results confirm that the
$\boldsymbol{\xi}_1$ is weakly polar, in accordance with its AFD-like
nature. In contrast, the second soft mode $\boldsymbol{\xi}_2$ is
found to have a considerably polar character. It can then be trivially
shown that the two soft modes have a very similar contribution to the
total spontaneous polarization of the $Cmc2_1$ phase of LTO, in spite
of the fact that $\boldsymbol{\xi}_1$ embodies 81\% of the PE-to-FE
distortion. Hence, the two-mode character of the transition is
critical to explain the relatively large $\boldsymbol{P}^{\rm S}$ of
the $Cmc2_1$ phase.

Unfortunately, we were unable to find experimental results for the
ferroelectric and response properties of the $Cmc2_1$ phase of
LTO. Table~III shows some results for the $P2_1$ phase of the compound
that is stable at room temperature. Interestingly, the dielectric and
piezoelectric responses measured for this phase seem compatible (at
least in magnitude) with our values for $Cmc2_1$. As regards the
spontaneous polarization, the value for the $P2_1$ phase is smaller
than ours by a factor of 6; this experimental result was essentially
confirmed by the first-principles study of Ref.~\onlinecite{bruyer10},
which employed a DFT theory similar to ours.\cite{fn:lda-vs-gga}
Hence, we can tentatively conclude that the $Cmc2_1$-to-$P2_1$
transformation, which is experimentally determined to occur at
1053~K,\cite{ishizawa82} involves a reduction of LTO's spontaneous
polarization. While such a reduction in the magnitude of
$\boldsymbol{P}^{\rm S}$ as the temperature decreases is not typical
in FE perovskites, in principle there is no reason to question this
possibility.

Finally, let us discuss an intriguing possibility suggested by the
very different magnitudes of the computed mode charges $\bar{Z}_{1,z}$
and $\bar{Z}_{2,z}$. Since $\bar{\boldsymbol{Z}}_s$ quantifies the
coupling between a polar mode and an applied electric field, we can
expect $\boldsymbol{\xi}_{2}$ to be much more reactive to an external
bias than $\boldsymbol{\xi}_{1}$. One can thus imagine the following
possibility: to apply an electric field to LTO in its $Cmc2_1$ phase
and switch {\em only the part of the spontaneous polarization
  associated to $\boldsymbol{\xi}_{2}$}. More precisely, if $(u^{\rm
  I}_1,u^{\rm I}_2)$ represents the FE phase discussed so far, an
electric field might allow us to take the material into a
qualitatively different polarization state $(u^{\rm II}_1,u^{\rm
  II}_2) \approx (u_1^{\rm I},-u_2^{\rm I})$. LTO would thus be a
four-state ferroelectric, as the $(-u^{\rm I}_1,-u^{\rm I}_2)$ and
$(-u^{\rm II}_1,-u^{\rm II}_2)$ variants would be accessible as
well. A necessary condition for such a partial switching to occur is
that the $(u^{\rm II}_1,u^{\rm II}_2)$ state be a minimum of the
energy. More precisely, we need the coupling between
$\boldsymbol{\xi}_{1}$ and $\boldsymbol{\xi}_{2}$ to be dominated by
the $\gamma'' u_1^2 u_2^2$ term of Eq.~(\ref{eq:E}): Note that, for
$|\gamma''|$ much greater than $|\gamma'|$ and $ |\gamma'''|$, the
four states $(\pm u^{\rm I}_1,\pm u^{\rm I}_2)$ would have essentially
the same energy; in contrast, if $\gamma'$ or $\gamma'''$
dominate, the ``$\boldsymbol{\xi}_{2}$-switched'' state would have a
much higher energy and might not be a minimum of $E(u_1,u_2)$. In
order to confirm or disprove such a partial switching, we considered
the equilibrium structure of the $Cmc2_1$ phase (i.e., the one
reported in Table~I) and generated configurations in which the
$\boldsymbol{\xi}_{2}$ distortion was inverted and given various
magnitudes; we relaxed such transformed structures and invariantly
obtained the original $Cmc2_1$ phase as final result. Hence, while we
did not explore the $E(u_1,u_2)$ energy landscape in detail, our
results clearly indicate there is no stable
``$\boldsymbol{\xi}_{2}$-switched'' state. Equivalently, this implies
that the $\gamma'$ and $\gamma'''$ terms of Eq.~(\ref{eq:E}) dominate
over $\gamma''$ (which supports the assumption made in Section~III.C).

\subsection{BaTiO$_3$-like ferroelectric modes in La$_2$Ti$_2$O$_7$}

Thus far we have discussed the main features of LTO's high-temperature FE
transition, showing that this compound is very different from the FE oxides
with the ideal perovskite structure. In particular, our results seem to
suggest that LTO and BTO have little in common, in spite of the fact that they
share the same building block: TiO$_6$ octahedra with Ti$^{4+}$ in the 3$d^0$
electronic configuration. In the following we show that such a conclusion
would be a deceptive one.

Ferroelectricity in the usual FE perovskite oxides relies on strong
dipole-dipole interactions whose fingerprint is the anomalously large
magnitude of the Born effective charges of some atomic
species.\cite{zhong94} A prototypical example of this behavior is BTO,
for which we computed the effective-charge tensors shown in
Table~V. Most notably, in the PE phase the Ti$^{4+}$ cations display
$Z^{*}$ values above 7$e$, almost doubling their nominal ionization
charge. Analogously, the displacement of the O atoms towards the Ti
has a dynamical charge of $-$5.77$e$ associated to it, almost tripling
the nominal value of $-$2$e$. It is well-known that this effect is
related with a partial covalent character of the Ti--O bond, and an
increased hybridization of the O-2$p$ and Ti-3$d$ orbitals as the Ti
and O atoms approach.\cite{posternak94} Thus, given this chemical
origin, the effective charges become less anomalous once the FE
distortion freezes in: according to the results in Table~V, Ti's
charge of 7.31$e$ gets reduced to 5.83$e$, while O's charge of
$-$5.77$e$ falls to $-$4.79$e$.

As apparent from Table~IV, some Ti and O atoms in LTO also present
anomalously large effective charges that reach values similar to those
obtained for BTO. In the case of the Ti atoms, the computed
$\boldsymbol{Z}^{*}$ tensors are rather isotropic, with the maximum
$Z^{*}$ values [i.e., 6.91$e$ for Ti(1) and 7.45$e$ for Ti(2)]
corresponding to displacements along the in-layer directions $x$ and
$z$. In the case of the O atoms, O(6) and O(7) clearly present the
largest values, which exceed $-$5$e$ as in BTO. Such giant dynamical
charges corresponds to displacements along the $x$ direction, for
which LTO presents infinite chains of TiO$_6$ octahedra (see Figs.~1
and~2) as those in the ideal perovskite structure. Hence, as regards
their {\em polarizability} properties, the results for the Ti and O
atoms in LTO are strongly reminiscent of the behavior that is
well-known for BTO.

Do such anomalous effective charges lead to FE instabilities in LTO?
In accordance with the above discussion, the answer to this question
is a negative one: the $Cmcm$ phase presents only two $\Gamma$-point
instabilities (i.e., the already discussed $\boldsymbol{\xi}_1$ and
$\boldsymbol{\xi}_2$) and the $Cmc2_1$ is stable against
$\Gamma$-point perturbations.\cite{fn:stableFE} Nevertheless, by
inspecting the $\boldsymbol{K}$-eigenmodes of the $Cmcm$ phase, it is
easy to find several low-energy FE modes that are BTO-like, i.e., they
involve the displacement of the Ti atoms away from the center of
nearly undistorted O$_6$ octahedra.\cite{fn:localdist,iniguez00} For
example, we found a marginally stable and strongly polar mode for
which we computed $\kappa_s =$~0.26~eV/\AA$^{2}$ and $\bar{Z}_{s,x}
=$~21.7$e$ (this distortion is $x$-polarized and has $B_{3u}$
symmetry); this mode gives an enormous contribution to the
$\epsilon_{xx}$ dielectric response:\cite{fn:modeepsilon} the obtained
value exceeds 600.  Interestingly, these BTO-like modes become stiffer
in the $Cmc2_1$ phase; thus, they do not give raise to any anomalously
large contribution to the dielectric response reported in Table~III.

In conclusion, our analysis shows that LTO presents obvious traces of
the FE instabilities of BTO. Such a {\em transferability} of
instabilities was demonstrated by one of us in an hexagonal polymorph
of BTO.\cite{iniguez00} In that case, the FE phase transition was
shown to be driven by soft modes that are an almost perfect match of
the usual BTO-like FE instability. In LTO, such modes are very low in
energy, but still stable; further, they clearly compete with the
dominating AFD-like instability, and become stiffer once the
$Cmcm$-to-$Cmc2_1$ transition occurs. Nevertheless, noting that
BTO-like FE distortions tend to be very sensitive to cell strains, our
results suggest the possibility that such instabilities might be
induced in LTO by suitable strain engineering or chemical
substitution, or that they might occur spontaneously in similar
layered perovskites.

\subsection{Implications for work on magnetoelectrics}

Finding materials that display large magnetoelectric (ME) effects
(i.e., a large magnetic reaction to an applied electric field) at room
temperature is a major challenge that remains to be successfully
tackled. The difficulties involved in the design of good
magnetoelectrics have been discussed elsewhere.\cite{wojdel10} At
present, the strategies that seem most promising rely on finding
systems that satisfy the following two conditions: (1) their atomic
structure must react strongly to an applied electric field (i.e., we
are looking for good dielectrics), and (2) the field-induced
distortions must have a large effect on the magnetic interactions (as
emphasized in Refs.~\onlinecite{fennie08} and \onlinecite{delaney09}).

While there are well-known strategies to comply with the first
condition,\cite{wojdel10} satisfying the second one is proving much
more difficult. In fact, the existing quantitative studies indicate
that the dielectric response of ME multiferroics like BiFeO$_3$ is
dominated by modes that have small magnetostructural couplings
associated to them.\cite{wojdel09} It is thus interesting to consider
the alternative approach adopted by Benedek and
Fennie:\cite{benedek10} These authors noted that modes involving O$_6$
rotations are likely to be strongly coupled with the magnetism of
perovskite oxides, as such AFD-like distortions usually control the
nature and magnitude of the main magnetic interactions (e.g., the
metal--oxygen--metal super-exchange and Dzyaloshinskii-Moriya
couplings). Hence, they looked for materials in which AFD-like
distortions are related with (or lead to) ferroelectricity, as in such
cases one could use an electric field to act upon the O$_6$
rotations. Note that the seeked connection between AFD-like
distortions and ferroelectricity is an exotic one, since the
O$_6$-rotational modes are strictly non-polar in compounds with the
ideal perovskite structure (see Section~III.B). Accordingly,
ferroelectricity-related O$_6$-rotational modes have been found in
non-ideal perovskites, e.g., in the PbTiO$_3$/SrTiO$_3$ artificial
superlattices\cite{bousquet08} and the layered compound
Ca$_3$Mn$_2$O$_7$.\cite{benedek10} In both cases, ferroelectricity has
an {\em improper} character, and a combination of various AFD-like
modes must occur for an spontaneous polarization to appear.

There is no need to emphasize the importance of our findings for LTO
in the context of magnetoelectrics. In LTO's case, the FE soft mode is
AFD-like already; hence, it can freeze in and give raise to a sizeable
spontaneous polarization by itself, without the need of any
accompanying distortion. Further, LTO's O$_6$ rotations couple
directly (bi-linearly) with an applied electric field, which might
prove advantageous for the purpose of obtaining large ME effects. (In
improper ferroelectrics, the coupling between an applied field and the
AFD-like modes will typically be a higher-order effect.)  Indeed, the
{\em proper ferroelectricity driven by O$_6$ rotations} that occurs in
LTO seems to be the dreamed FE instability from the viewpoint of ME
applications.

One would thus like to obtain LTO-like ferroelectricity in a magnetic
material. The most obvious possibility is to consider the substitution
of Ti by Mn in LTO, so as to form La$_2$Mn$_2$O$_7$, a crystal that we
have not found described in the literature. In La$_2$Mn$_2$O$_7$ we
would have manganese in the Mn$^{4+}$ ionization state, most likely in
the high spin $t_{2g}^{3}e_{g}^{0}$ electronic configuration; thus, we
can expect this crystal to be insulating. Further, since the Ti--O
chemistry does not seem to play any relevant role in LTO's FE
instability, we can expect La$_2$Mn$_2$O$_7$ to present a FE
transition analogous to LTO's. Hence, this material seems an excellent
candidate to satisfy the condition (2) mentioned above and thus
display large ME effects. Additionally, one would like the FE
transition to occur near room temperature, so as to benefit from the
enhancement of the functional responses near $T_{\rm C}$ [in the
spirit of the condition (1) mentioned above]. In this sense, it may be
useful to note that materials like Sr$_2$Ta$_2$O$_7$\cite{ishizawa76}
or Sr$_2$Nb$_2$O$_7$\cite{ishizawa75} present FE transitions that seem
similar to LTO's but occur at lower $T_{\rm C}$'s: 166~K and 1615~K,
respectively. This suggests that exploring alternative compositions is
a promising route to tune the temperature of the O$_6$-rotational FE
transition.

\section{Summary and Conclusions}

We used first-principles methods to study the origin of
ferroelectricity in the layered perovskite La$_2$Ti$_2$O$_7$ (LTO),
one of the materials with highest Curie temperature known ($T_{\rm
  C}=$~1770~K). To do so, we carried out for LTO a research program
that has been repeatedly and successfully applied to the investigation
of displacive phase transitions in ferroelectric (FE) oxides with the
perovskite structure. Our results allowed us to characterize LTO's
high-temperature FE transition, which was found to present very
peculiar features.

We found that ferroelectricity in LTO is very different from what
occurs in the well-known FE oxides with the ideal perovskite
structure, such as BaTiO$_3$ (BTO) or PbZr$_{1-x}$Ti$_x$O$_3$
(PZT). Indeed, the dominant FE instability of this compound has little
in common with the text-book picture of positive charges moving
against negative charges in an ionic insulator; instead, it involves
concerted rotations of the oxygen octahedra that form the perovskite
framework. Hence, LTO's high-temperature structural transition is
reminiscent of the behavior of simple perovskite oxides, but not the
FE ones: LTO's FE distortion is much alike the O$_6$ rotational soft
modes that drive the {\em anti-ferrodistortive} phase transitions of
SrTiO$_3$, LaAlO$_3$, LaMnO$_3$, and many other {\em non-polar}
perovskite crystals.

Hence, we found that the existence of ferroelectricity in LTO at
record-high temperatures is the result of structural distortions as
those occurring in many perovskite oxides that are {\em not}
ferroelectric. As discussed here, the solution to this puzzle has to
do with LTO's layered structure: Because the lattice of oxygen
octahedra is truncated in this compound, the O$_6$-rotations acquire a
polar character and give raise to a macroscopic polarization. We can
thus describe LTO as a {\em topological ferroelectric}, since it owes
its spontaneous polarization to the layered topology of its
structure. Note that such a peculiar topology does not seem essential
for LTO's high-temperature transition to occur, but it is critical for
it to have a ferroelectric character.

We quantified energetics of LTO's FE transformation, the results being
consistent with the very high temperature at which it happens. Most
remarkably, we found that the structural distortion connecting the PE
($Cmcm$) and FE ($Cmc2_1$) phases presents large contributions from
two modes, namely, the above-mentioned strong instability consisting
of O$_6$ rotations, and an isosymmetric weakly-unstable mode that
involves deformations of the oxygen octahedra. Further, we found that
the large energy change associated to the $Cmcm$-to-$Cmc2_1$
transition depends crucially on the simultaneous occurrence of both
modes. We were able to describe such effects in terms of a very strong
and {\em cooperative} anharmonic coupling, and briefly discussed the
possibility of constructing a phenomenological theory of such a {\em
  two-mode transition}. Interestingly, LTO's behavior is in strong
contrast with what (to the best of our knowledge) is most common among
perovskite oxides, where structural instabilities tend to compete and
suppress each other.

We calculated the polarization and response properties of LTO's FE
phase, obtaining results that are consistent with related experimental
information. Interestingly, the computed electric properties revealed
clear similarities between LTO and prototype ferroelectric BTO. For
example, we obtained anomalously large Born effective charges for some
Ti and O atoms in LTO, a feature that is known to be the fingerprint
of the FE instabilities in compounds like BTO and PZT. Further, our
results showed that LTO presents nearly-unstable modes that are
strongly polar and involve atomic distortions that essentially match
BTO's FE instabilities. Such traces of BTO-like behavior in LTO
suggest the intriguing possibility that {\em conventional}
ferroelectricity might be induced in this compound upon suitable
modifications (e.g., strain engineering or chemical substitutions), or
that such a behavior might occur spontaneously in similar materials.

Finally, let us emphasize the implications that our findings have for
the design of new magnetoelectric multiferroics. In the context of
magnetoelectric perovskite oxides, it would be ideal to have proper
ferroelectricity driven by O$_6$-rotational modes, so that an electric
field can be used to tune the structural distortions (e.g.,
metal--oxygen--metal angles) that control the main magnetic
interactions (i.e., super-exchange and Dzyaloshinskii-Moriya). That is
exactly the kind of ferroelectricity that we have found in LTO. Here
we have briefly discussed how to obtain such an effect in a magnetic
perovskite, proposing La$_2$Mn$_2$O$_7$ as the most promising
candidtate material.

In conclusion, our theoretical study of La$_2$Ti$_2$O$_7$ has revealed
a wealth of interesting effects, some of which may have important
implications for current work on multifunctional oxides. We thus hope
this study will stimulate further investigation of these layered
perovskites and the novel functionalities that they may offer.

\vskip 0.5mm

This work was supported by MICINN-Spain (Grants No. MAT2010-18113,
No. MAT2010-10093-E, and No. CSD2007-00041) and CSIC's JAE-pre program
(J.L.P.). We used the supercomputing facilities provided by CESGA, the
tools provided by the Bilbao Crystallographic Server,\cite{bilbao} and
the {\sc vesta} software\cite{vesta} for the preparation of some
figures.  We gratefully acknowledge discussions with E.~Canadell,
M. Stengel, and J.C. Wojde\l.


\begin{thebibliography}{99}
\bibitem{fn:highTphase} In several publications it is assumed that the
  high-temperature phase of LTO presents the $Cmcm$ space group, and
  the literature is cited for supposed experimental proof of this
  fact. However, we have carefully checked the literature and were
  unable to find any results concerning the crystal structure of LTO's
  PE phase.
\bibitem{fn:unbounded} The fourth-order expression in Eq.~(2) may not
  be good description of the energy for arbitrary values of $u_1$ and
  $u_2$; e.g., distortions with $u_1 = u_2$ are unbounded from below
  for the fitted parameter values. An acceptable description of the
  full $E(u_{1},u_{2})$ energy landscape would probably require including
  higher order terms.
\bibitem{fn:c} The $Q_1^3Q_2$ term guarantees that both $Q_1$ and
  $Q_2$ will freeze in at $T_{\rm C}$.
\bibitem{fn:stableFE} The $Cmc2_{1}$ phase might present structural
  instabilities associated to the $Cmc2_1$-to-$P2_1$ transition that
  is known to occur in LTO. Such a transition involves the doubling of
  the $Cmc2_1$ cell along the $x$ direction. Thus, the
  $Cmc2_1$-to-$P2_1$ transformation is driven by a non-$\Gamma$
  distortion; such distortions were not considered in the present
  study.
\bibitem{fn:localdist} To quantify the resemblance between the TiO$_6$
  distortions in BTO and LTO, we projected the corresponding atomic
  displacements of the $\boldsymbol{K}$-eigenmodes of LTO's $Cmcm$
  phase into the distortion pattern derived form the FE soft mode of
  BTO's $Pm\bar{3}m$ phase. This kind of analysis is described in more
  detail in Ref.~\protect\onlinecite{iniguez00}.
\bibitem{fn:modeepsilon} The lattice dielectric response can be split
  in mode contributions as $\epsilon_{\alpha\beta} = \Omega^{-1}
  \sum_s \bar{Z}_{s,\alpha} \bar{Z}_{s,\beta} \kappa_{s}^{-1}$, in
  units of $\epsilon_0$ and where $\Omega$ is the cell volume.
\end{thebibliography}
\end{document}